\documentclass[12pt]{article}
\usepackage{graphicx}
\usepackage{url}
\usepackage{threeparttable}
\usepackage{pdflscape}
\usepackage{xcolor}

\usepackage{amssymb}
\usepackage{bbm}
\usepackage{marginnote}
\usepackage{amsmath}
\usepackage{xcolor}
\usepackage{lipsum}

\usepackage{amssymb, color}
\usepackage{bbm}
\usepackage{latexsym}
\usepackage{mathrsfs}
\usepackage{amsmath}
\usepackage{lscape}
\usepackage{amsfonts}
\usepackage{indentfirst}
\usepackage{graphics}
\usepackage{bm}
\usepackage{natbib}
\usepackage{epsfig}
\usepackage[normalem]{ulem}
\usepackage{multirow}
 \usepackage{float}

\usepackage{bm}
\def\bbeta{{\bm\beta}}
\def\hbbeta{{\bm{\hat \beta}}}
\def\bx{{\bm x}}

\def\bX{{\bm X}}

\def\bb0{{\bm 0}}
\def\bw{{\bm w}}

\def\be{{\bm e}}

\def\bM{{\bm M}}

\newcommand{\esp}{\end{sloppypar}}
\newcommand{\ee}{\end{equation}}
\newcommand{\beanno}{\begin{eqnarray*}}
\begin{document}
\parindent 4mm
\baselineskip 20pt

\title{Efficient Computational Algorithm for Optimal Continuous Experimental Designs}

\author{\normalsize Jiangtao Duan$^1$\qquad Wei Gao$^{1}$ \qquad Hon Keung Tony Ng$^{2}$
\\[1ex]
\normalsize
$^1 \star$Key Laboratory for Applied Statistics of MOE, School of Mathematics and Statistics, \\
\normalsize
Northeast Normal University, Changchun, Jilin 130024, China\\
\normalsize
$^{ 2}$Department of Statistical Science, Southern Methodist University, Dallas, TX 75275, USA
}

\maketitle

\begin{abstract}
A simple yet efficient computational algorithm for computing the continuous optimal experimental design for linear models is proposed. An alternative proof the monotonic convergence for $D$-optimal criterion on continuous design spaces are provided. We further show that the proposed algorithm converges to the $D$-optimal design. We also provide an algorithm for the $A$-optimality and conjecture that the algorithm convergence monotonically on continuous design spaces. Different numerical examples are used to demonstrated the usefulness and performance of the proposed algorithms.
\end{abstract}

\noindent {\bf Key Words:} {Continuous experimental design; $D$-optimal, $A$-optimal; Regression model}

\section{Introduction}

In science, engineering and social science, experiments are frequently used to gain the knowledge about relationships between independent and dependent variables. A well-designed experiment can maximize the amount of information that can be obtained from the experiment. The field of design of experiments deals with statistical approaches for efficient experimentation, i.e., deriving the required information about at the least expenditure of resources (Barker, 1994).

As pointed out by Vanhatalo, V\"{a}nnman and Hyllander (2007) and Vanhatalo and Bergquist (2007), discontinuous processes, i.e., processes where parts or batches are produced, dominate the applications of design of experiments in practice as well as in the literature. However, there are many situations that measurements are taken on a continuous process in different industries (e.g., chemical or steel production, ore processing, etc). In continuous processes the product gradually and with minimal interruptions passes through a series of different operation and the product exhibits characteristics such as liquids, powders, slurries, and pellets (Fransoo and Rutten, 1994 and Vanhatalo, et al., 2007). For instance, mineral ore is processed and handled in a continuous stream, and  steel is continuously cast and rolled into sheet metal. Design of experiment with continuous process is of a great interest.


Consider an experiment consisting of a set of independent trials. Assume that the observed response of each trial depends on a design point chosen from a finite set $\mathcal{X}$ representing those possible experimental conditions. For $\bw \in \mathcal{X}$, the real-valued observation $y(\bw)$ is assumed to satisfy the linear regression model
$$y(\bw)=\bX^{T}(\bw)\bbeta+\varepsilon(\bw),$$
where $\bX(\bw)\in \mathbb{R}^{m}$ is the regressor (independent variable) associated with $\bw$, the vector $\bbeta \in \mathbb{R}^{m}$ contains the unknown parameters of the model, and $E[\varepsilon(\bw)] = 0$, $Var[\varepsilon(\bw)]=\sigma^{2}$, with $\sigma^{2}>0$, is an unknown quantity. For different independent trials, the errors are assumed to be uncorrelated and independent. We assume that the model be non-singular in the sense that $\{\bX(\bw):\bw\in \mathcal{X}\}$ spans $\mathbb{R}^{m}$.
Under this model and those assumptions, an exact design of experiment of size $N$ is a selection $\xi$ of $m$ design points, $\bw_{1},\bw_{2},\cdots,\bw_{m}$, and $n_{1}, n_{2}, \cdots, n_{m} (m \neq 1)$, where $n_{k}$ ($k = 1,2, \ldots, m)$ denotes the number of times the design point $\bw_{k}$ occurs the $m$ design points with $\bw_{k} \in \mathcal{X}$,  $N=n_{1}+\cdots+n_{m}$.
This experimental design could be summarized by defining a design $\xi_{N}$ as
$${\xi}_{N} = \left(\begin{array}{ccc}\bw_{1} & \cdots & \bw_{m} \\ \frac{n_1}{N} & \cdots & \frac{n_{m}}{N}\end{array}\right).$$
We can show that an exact experimental design $\xi$ of size $N$ can be represented by a probability distribution on $\mathcal{X}$ in which the probability is $n_{j}/N$ when $\xi_{N}$ occurs at $\bw_{j}$.

If $\xi$ follows a continuous probability distribution or a discrete probability distribution, then we have
$$
{\xi} = \left(\begin{array}{ccc}\bw_{1} & \cdots & \bw_{m} \\ \eta_{1} & \cdots & \eta_{m}\end{array}\right),
$$
where $\Pr(\xi=\bw_{j})=\eta_{j}, j=1,\cdots,m$ and $\sum_{j=1}^{m} \eta_{j} = 1$, and $\xi$ is continuous design. 
For a complete overview on the subject of the theory of opetimal design of experiments, one can refer to Dette and Studden (1997) and the references therein.


We denote the information matrix of $\xi$ for the exact experimental design by
$$
\bM(\xi)=\sum\limits_{i=1}^{m}\mu_{i}\bX(w_{i})\bX^{T}(w_{i}),
$$
where $\mu_{i}=n_{i}/N$ is the weight corresponding to the point $\bw_{i}$.
\par
Similarly, for the continuous experimental design, we have the information matrix
$$
\bM(\xi)=\int_E f(w)\bX(w)\bX^{T}(w)d\mu(w).
$$
Based on this formulation, Chen (2003) provided an approximate $D$-optimal design  for quadratic polynomial when the design space is circle. Under the model assumptions, the information matrix is proportional to the Fisher information matrix corresponding to $\bbeta$. Therefore, the general aim is to choose $\xi$ such that $\bM(\xi)$ is as large as possible.

There is a rich literature on the theory of optimal design and different numerical computational algorithms have been proposed to obtain optimal designs under different scenarios. For exact experimental designs, the enumeration techniques based on the branch-and-bound principle are able to solve medium-sized problems (see, for example, Welch, 1982 and Uci{\'n}ski and Patan, 2007), and the stochastic optimization methods also used to obtain the optimal designs (e.g., Haines, 1987). Harman and Filov\'{a} (2014) considered a $DQ$-optimality, which is a quadratic approximation of the $D$-optimality criterion in the neighborhood of the approximate $D$-optimal information matrix, to compute exact experimental designs for linear regression models. Gao, et al. (2014) develop efficient computational algorithms to obtain optimal allocation for a general regression model subject to the $D$-optimality and $A$-optimality criteria. There are many analytic methods for constructing approximate optimal designs to instead of continuous optimal designs. Fedorov (1972) and Wynn (1970) developed the so-called vertex direction methods (VDMs). In each iteration, VDMs move the current design $\xi$ in the direction of $\be_{\bw}$ for some design point $\bw$ while decreasing all remaining components of $\xi$ by the same proportion.  Silvey, Titterington and Torsney (1978) proposed a multiplicative algorithm, which are simple iterative strategies that converge monotonically, i.e., the determinant criterion never decreases along woth the iterations. Yu (2011) proposed a new algorithm that extends the VDM and multiplicative algorithm. Although these existing methods converge and some of them converge monotonically, these algorithms tend to be slow. Recently, Castro, et al. (2018) use the moment-sum-of-squares hierarchy of semi-definite programming problems to solve the approximate optimal design problem numerically. Harman, Filov{\'a} and Richt{\'a}rik (2018) proposed the randomized exchange method (REX), which is a simple batch-randomized exchange algorithm, for the optimal design problem. The REX method can be viewed as an efficient extension or combination of both the VEM algorithm and the $KL$-exchange algorithm that is used to compute exact designs (Vanhatalo, et al., 2007).

In this paper, we study the numerical computation of the optimal continuous designs directly, with special focus on the $D$-optimality and $A$-optimality criteria. We propose a simple computational algorithm to obtain the continuous experimental designs. Mathematical results related to the convergence and monotonicity of the proposed algorithms are developed. A simulation study is performed to show the reliability of these algorithms. The paper is organized as follows. In Section 2, we consider the inference based on a linear regression model and present the form of the information matrix and the variance-covariance matrix when the design is continuous. The proposed computational algorithms for the two commonly used optimal criteria, $D$-optimality and $A$-optimality criteria, and their related mathematical properties are discussed in Section 2. Section 3 presents some numerical illustrations with several linear regression models for continuous $D$-optimality and $A$-optimality designs. The proofs of the main results are presented in the Appendix.

\section{Algorithms for Optimal Designs}

Consider the linear regression model
\begin{equation}
y(w)=\bx^{T}(w){\bbeta}+\epsilon(w),\;\;w\in E,
\label{MD}
\end{equation}
where $\bbeta$ is a $p$-dimensional parameter vector, $\bx(w)$ is the covariates, $E$ is the continuous design space and $\epsilon(w)$ is error terms with mean $0$ and variance $\sigma^2$. When the observations are obtained based on Model (\ref{MD}), the ordinary least square estimator of $\bbeta$ can be expressed as
$$
{\hbbeta}=\left[\int_E f(w)\bx(w)\bx^{T}(w)d\mu(w)\right]^{-1}\int_E y(w)x(w)f(w)d\mu(w),
$$
where $f(w)\geq 0$ is the mass on the point $w$ and
$$
\int_E f(w)d\mu(w)=1.
$$
The variance of $\hbbeta$ is
$$
Var(\hbbeta)=\left[\int_E f(w)\bx(w)\bx^{T}(w)d\mu(w)\right]^{-1}.
$$
Most of the existing computational algorithms for obtaining the optimal designs set a finite design space $\mathcal{X}=\{x_{1}, \cdots, x_{n}\} \subset \mathbf{R}^{m}$, which utilize the idea of discretizing the underlying continuous space. Here, we proposed algorithms that directly obtain the optimal design of continuous space for for $D$-optimality  and $A$-optimality without the discretization of the continuous space.

\subsection{Algorithm for $D$-optimal designs}

In an experiment, researchers often wish to estimate the model parameters with maximum precision and minimum variability possible. An optimality criterion often used in experimental design is based on the determinant
of the Fisher information matrix, which equals to the reciprocal of the determinant of the asymptotic variance-covariance matrix. The $D$-optimal design maximizes the log-determinant of the information matrix, i.e., it minimizes the log-determinant of the asymptotic variance-covariance matrix $Var(\hat\bbeta)$. The $D$-optimal criterion can be described as follows.
\par
\textbf{$D$-optimal criterion:}
\begin{equation}
\min\limits_{f(w)}\left\{-\log|\int_E f(w)\bx(w)\bx^{T}(w)dw|:\;\mbox{subject to}\;f(w)\geq 0\;\mbox{and}\;\int_E f(w) dw=1\right\},
\label{DO}
\end{equation}
We can obtain the following result.

{\bf Theorem 1.} $f^{*}(w)$ is the $D$-optimal solution for (\ref{DO}) if and only if
$$
\int_E f(w)\bx^{T}(w)\left[\int_E f^{*}(u)\bx(u)\bx^{T}(u)du\right]^{-1}\bx(w)dw\leq p
$$
for $f(w)\geq0$ and $\int_E f(w)dw = 1$.

The proof of Theorem 1 is provided in Appendix.
Based on Theorem 1, for the $D$-optimal criterion in ($\ref{DO}$), the following algorithm is proposed to obtain the optimal $f^{*}(w)$.

{\bf Proposed algorithm for $D$-optimal design:}


Step D1. Set the initial value of $f$ as

$f^{(0)}(w) =
\begin{cases}
1 & {\mbox {for }} w \in E, \\
0 & {\mbox {otherwise.}}
\end{cases}$

Step D2. Update $f$ in the $n$-th step as

$$
f^{(n)}(w)=\frac{f^{(n-1)}(w)\bx^{T}(w)D^{(n-1)}\bx(w)}{p},
$$
where
$$
D^{(n-1)}=\left[\int_E f^{(n-1)}(w)\bx(w)\bx^{T}(w)dw\right]^{-1}.
$$

Step D3. Repeat Steps D1 and D2 until convergence occurs.

Here, we provide a proof of the convergence of the proposed algorithm for $D$-optimality. To prove the convergence of the algorithm, we first prove that the log-determinant of the information matrix is monotonous. In order to prove its convergence, we must add the bounded assumption and require the following two lemmas.

{\bf Lemma 1.} Let $A(w)$ be a nonnegative definite matrix function on $E$ , $f(w)$ and $g(w)$ are nonnegative functions on it, and
$$\int_E f(w) A(w) dw\;\;\mbox{and}\;\;\int_E g(w) A(w) dw$$
are positive definite matrices with infinite elements. Then,
\begin{eqnarray*}
&      & \log|\int_E f(w)A(w)dw|-\log |\int_E g(w) A(w)dw| \\
& \geq & \int_E g(w)\mbox{tr}\left\{A(w)[\int_E g(u)A(u)du ]^{-1}\right\}\log\frac{f(w)}{g(w)}dw.
\end{eqnarray*}
\textbf{Proof.} Following the proof of Lemma 2 in Gao, Chan, Ng and Lu (2014), Lemma 1 can be proved. 

{\bf Lemma 2.} Suppose $f(w)$ and $g(w)$ are density functions defined on $E$, then
$$
\int_E|f(w)-g(w)|dw\leq\left\{2\int_E f(w)\log\frac{f(w)}{g(w)}dw\right\}^{1/2}.
$$
\textbf{Proof.} See Kullback (1967). 

Now, we present the theorem to show the convergence of the proposed algorithm for $D$-optimality.

{\bf Theorem 2.} Under the assumption that $\log |\int_E \bx(w)\bx^{T}(w)dw|$ is bounded, we have
$$
\int_E |f^{(n)}(w)-f^{(n-1)}(w)|dw\longrightarrow 0 {\mbox { as }} n \longrightarrow +\infty.
$$

The proof of Theorem 2 is provided in Appendix.
The algorithm proposed here is in the general class of the multiplicative algorithms (Silvery, et al., 1978), which shares the simplicity and monotonic convergence property of class of the multiplicative algorithm. One of the most attractive properties of the proposed algorithm is that the convergence rate of the algorithm does not depend on the number of design points $N$ compared to some existing algorithms such as the  coordinate-exchange algorithm (Meyer and Nachtsheim, 1995) and the randomized exchange algorithm (Harman and Filov{\'a}, 2014).

\subsection{Algorithm for $A$-optimal designs}

Another optimality criterion considered here is based on the trace of the first-order approximation of the variance-covariance matrix of the MLEs. It is identical to the sum of the diagonal elements of the variance-covariance matrix. The $A$-optimality criterion provides an overall measure of the average variance of the parameter estimates. The $A$-optimal design points minimize the objective function defined by
\textbf{$A$-optimal criterion:}
\begin{equation}
\min\limits_{f(w)} \left\{ {\mbox {tr}} \left( \left[\int_E f(w)\bx(w)\bx^{T}(w)dw \right]^{-1} \right), \mbox{subject to}\;f(w)\geq 0\;\mbox{and}\;\int_E f(w) dw=1\right\}.
\label{AOP}
\end{equation}

{\bf Theorem 3.} $f^{*}(w)$ is the $A$-optimal solution for (\ref{AOP}) if and only if
$$
\frac{ {\mbox {tr}}(\left[I_E(f^{*}, \bx) \right]^{-1} I_E(f, \bx) \left[ I_E(f^{*}, \bx) \right]^{-1})} {{\mbox {tr}}(\left[I_E(f^{*}, \bx)\right]^{-1})} \leq 1,
$$
for $ f(w)\geq0$ and $\int_E f(w)dw=1$.

The proof of Theorem 3 is provided in Appendix. Based on Theorem 3, for the $A$-optimal criteria in (\ref{AOP}), the following algorithm is proposed to obtain the optimal design $f^{*}(w)$.

{\bf Proposed algorithm for $A$-optimal design:}


Step A1. Set the initial value of $f$ as
$f^{(0)}(w) =
\begin{cases}
1, & {\mbox {for }} w \in E, \\
0, & {\mbox {otherwise.}}
\end{cases}$


Step A2. Update $f$ in the $n$-th step as
$$
f^{(n)}(w)=\frac{f^{(n-1)}(w)}{p} \left[(p-1)\frac{ {\mbox {tr}}(D^{(n-1)}\bx(w)\bx^{T}(w)D^{(n-1)})}{{\mbox {tr}}(D^{(n-1)})}+1 \right],
$$
where
$$
D^{(n-1)}=\left[\int_E f^{(n-1)}(w)\bx(w)\bx^{T}(w)dw\right]^{-1}.
$$

Step A3. Repeat Steps A1 and A2 until convergence occurs.

For $A$-optimality, the proposed algorithm does not depend on the initial value and it provides a convergence solution which is robust to the initial value. Although a theoretical justification of convergence of the proposed computational algorithm for $A$-optimality is not available, simulation results strongly support the validity and reliability of the algorithm. An extensive simulation study presented in subsequent section is used to study the properties of the algorithm for $A$-optimality. Based on the simulation results, we conjecture the monotonic convergence of the algorithm for $A$-optimality.

\section{Numerical Illustrations}

To further illustrate the proposed algorithms can efficiently identify the optimal design, we consider four different settings to study the convergence of the proposed algorithms.

{\bf Setting 1:} Consider the model
$$y(w)=\bbeta^{T}\bx(w),$$
where
\begin{eqnarray*}
& & \bbeta = (\beta_1,\beta_2)^{T}, \\
& & \bx(w) = (w_1,w_2)^{T}, \\
& & w \in E=\{(w_1,w_2)^{T}: \;w_1^2+w_2^2\leq 1\}.
\end{eqnarray*}

{\bf Setting 2:}
$$
y(w)=\bbeta^{T}\bx(w),
$$
where
\begin{eqnarray*}
& & \bbeta=(\beta_0,\beta_1,\beta_2,\beta_3,\beta_4,\beta_5)^{T},\\
& & \bx(w)=(1,w_1,w_2,w_1^2,w_1w_2,w_2^2)^{T},\\
& & w\in E=\{(w_1,w_2)^{T}: \;w_1^2+w_2^2\leq 1\}.
\end{eqnarray*}

In Settings 1 and 2, we consider the continuous design space $E = \{(w_1,w_2)^{T}: \;w_1^2+w_2^2\leq 1\}$ and polynomial measurements with unknown parameters $\bbeta \in \mathbb{R}^{d}$, where the number of unknown parameters of Settings 1 and 2 are $d_{1} = 2$ and $d_{2} = 6$, respectively. To compute the $D$-optimal design and the $A$-optimal design, we apply the algorithms presented in Sections 2.1 and 2.2. We can obtain the positions of the optimal design points. For Settings 1 and 2, Figures 1--4 show that the optimal design points converge to the boundary of a circle and the support also includes the point $(0, 0)$ if the model includes the constant term. For Setting 2 (see Figure 3), we integrate the blue region around (0, 0) and get the integral value of 0.162 ($\approx 1/6$), and the integral of area of the circle is 0.833 ($\approx 5/6$).
Note that 1/6 and 5/6 are the theoretical proportions for the optimal design in Setting 2. In order to verify that the optimal design points have uniform distribution on the ring, we also integrate region with the same arc on the ring and find that these integral values are equal. Similar conclusions can be obtained for $A$-optimal design presented in Figure 4.

We also compare the performance of our algorithm and the randomized exchange algorithm (Harman, et al., 2018) for computing the $D$-optimal design experiments and $A$-optimal design experiments for Setting 2, and the results are presented in the Tables \ref{DS2} and \ref{AS2}. From Tables \ref{DS2} and \ref{AS2}, we observe that the optimal design points indeed at the edge of the circle, but the weights of the randomized exchange algorithm do not seem to be equal to the theoretical value.

{\bf Setting 3:} Consider the model
$$y(w) = \bbeta^{T}\bx(w),$$
where
\begin{eqnarray*}
& & \bbeta=(\beta_0,\beta_1,\beta_2,\beta_3,\beta_4,\beta_5)^{T},\\
& & \bx(w)=(1,w_1,w_2,w_1^2,w_1w_2,w_2^2)^{T},\\
& & w\in E=[-1,1]\times[-1,1].
\end{eqnarray*}

For Setting 3, Figures 5 and 6 show that there are 9 optimal design points in the continuous design space $E$, and the weights are obtained by integrating around these 9 points. The weights obtained from the randomized exchange algorithm and the proposed algorithm are presented in Tables \ref{DS3} and \ref{AS3}. From Tables \ref{DS3} and \ref{AS3}, we can see that the proposed method identified the same optimal design points as the randomized exchange algorithm, and the corresponding weights are close to the theoretical values.

{\bf Setting 4:} Consider the model
$$y(w)=\bbeta^{T}\bx(w),$$
where
\begin{eqnarray*}
& & \bbeta=(\beta_0,\beta_1,\beta_2,\beta_3,\beta_4,\beta_5,\beta_6,\beta_7,\beta_8,\beta_9)^{T},\\
& & \bx(w)=(1,w_1,w_2,w_3,w_1^2,w_1w_2,w_1w_3,w_2^2,w_2w_3,w_3^2)^{T},\\
& & w\in E=[-1,1]\times[-1,1]\times[-1,1].
\end{eqnarray*}

In Setting 4, we consider the model in which the continuous design space is a three-dimensional cube. Using the proposed algorithm, we can obtain 27 optimal design points, and the weights are the integral around these points. For comparative purposes, we also compute the weights of those 27 optimal design points and compare with the optimal design obtained from the randomized exchange algorithm. These results are presented in Tables \ref{DS4} and \ref{AS4}. We can observe that the results obtained from the proposed algorithm are very close to the theoretical value. Note that in the results presented in Tables \ref{DS4} and \ref{AS4}, the optimal design points obtained by REX method may be unstable in the sense that the optimal design points may not be unique in multiple runs of the algorithm. For illustrative purpose, we take the optimal points based on the REX algorithm which are closest to the theoretical value and present the results in Tables \ref{DS4} and \ref{AS4}.

Based on the numerical evaluations of the four settings considered here, we found that the proposed algorithms for $A$-optimality and $D$-optimality converge in all cases, and the optimal design points as well as the corresponding weights are very close to the theoretical values. As the dimension increases, most existing computational algorithms need more design points to support the initial design, and more spaces are needed for the computer to to store these points, which will cause the program to run slowly. Moreover, we notice that the weight of each optimal design point is unstable using the REX algorithm. For example, in Settings 2 and 4, when the REX algorithm is run multiple times with different values of the number of design point $N$, the location of each optimal design point is not exactly the same and the corresponding weight is not exactly equal. This will bring uncertainty to the choice of optimal design points and their corresponding weights.


Although we focus on $D$-optimal designs and $A$-optimal designs for linear models, the basic ideas of the proposed algorithms are not limited to $D$-optimality and $A$-optimality, and the continuous design space can be extended to high-dimensional situations.

\section{Concluding Remarks}

In this paper, we proposed simple yet efficient iterative computational algorithms for obtaining the $D$-optimal and $A$-optimal continuous designs for linear models. We also provided an alternate proof of the monotonic convergence of the proposed algorithm for $D$-optimality and demonstrate that the proposed algorithm converges to the optimal design. Although a theoretical justification for the convergence of the proposed computational algorithm for $A$-optimality is not available, our computational results support the validity and reliability of the algorithm. These algorithms are programmed under MATLAB R2016a environment and the programs are available from the authors upon request.


\section*{Appendix}

\subsection*{Proof of Theorem 1}

Since $\log(|\textbf{A}|)$ is convex in $\textbf{A}$ with $\textbf{A}$ being a positive definite matrix, we have
\begin{eqnarray*}
&&\log \left\{ \left| (1-\lambda)\int_E f^{*}(w)\bx(w)\bx^{T}(w)dw+\lambda\int_E f(w)\bx(w)\bx^{T}(w)dw \right| \right\}\\
&&\geq (1-\lambda) \log\left\{\left|\int_E f^{*}(w)\bx(w)\bx^{T}(w)dw \right| \right\}+\lambda \log\left\{ \left|\int_E f(w)\bx(w)\bx^{T}(w)dw \right| \right\}.
\end{eqnarray*}
Then, $f^{*}$ is the optimal solution for the $D$-optimal criterion in (2) if and only if
\begin{eqnarray*}
\frac{\log \left\{ \left| (1-\lambda) I_{E}(f^{*}, \bx) + \lambda  I_{E}(f, \bx) \right| \right\} - \log \left\{ \left| I_{E}(f^{*}, \bx) \right\| \right\}}{\lambda} \leq 0
\end{eqnarray*}
where $I_{E}(f, \bx) = \int_E f(w)\bx(w)\bx^{T}(w)dw$, for all $f(w)$ that satisfy $f(w)\geq 0$ and $\int_E f(w) dw=1$, and $\lambda > 0$.
Thus, for $\lambda\downarrow 0$, we have
\begin{eqnarray*}
&   &\lim\limits_{\lambda\downarrow 0} \frac{\log(|(1-\lambda) I_{E}(f^{*}, \bx) + \lambda I_{E}(f, \bx)|) - \log(|I_{E}(f^{*},\bx)|)}{\lambda}\\
& = & \left. \frac{\partial \log \{ |(1-\lambda) I_{E}(f^{*}, \bx) +\lambda I_{E}(f, \bx)| \} }{\partial \lambda} \right|_{\lambda=0}\\
& = & \mbox{tr} \left\{ \left[\int_E f^{*}(w)\bx(w)\bx^{T}(w)dw \right]^{-1} \int_E [f(w)-f^{*}(w)]\bx(w)\bx^{T}(w)dw \right\}\\
& = & \mbox{tr} \left\{ \left[\int_E f^{*}(w)\bx(w)\bx^{T}(w)dw \right]^{-1} \int_E f(w)\bx(w)\bx^{T}(w)dw \right\} - p \leq 0,
\end{eqnarray*}
which gives the result stated in the theorem. 

\subsection*{Proof of Theorem 2}

From Lemma 1, we have
\begin{eqnarray*}
& & \log \left|\int_E f^{(n)}(w)\bx(w)\bx^{T}(w)dw \right|-\log \left|\int_E f^{(n-1)}(w)\bx(w)\bx^{T}(w)dw \right|\\
& \geq & \int_E f^{(n-1)}(w)\mbox{trace}\left\{\bx(w)\bx^{T}(w)D^{(n-1)}\right\}\log \frac{f^{(n)}(w)}{f^{(n-1)}(w)}dw\\
& = & \int_E f^{(n-1)}(w)\bx^{T}(w)D^{(n-1)}\bx(w)\log \frac{f^{(n)}(w)}{f^{(n-1)}(w)}dw\\
& = & p\int_E f^{(n)}(w)\log \frac{f^{(n)}(w)}{f^{(n-1)}(w)}dw\geq 0.
\end{eqnarray*}
Thus, we can conclude that
$$\log \left|\int_E f^{(n)}(w)\bx(w)\bx^{T}(w)dw \right|$$
is increasing in $n = 2, 3, \ldots.$ We can get that
\begin{eqnarray*}
\log \left|\int_E f^{(n)}(w)\bx(w)\bx^{T}(w)dw \right| \leq \log \left|\int_E \bx(w)\bx^{T}(w)dw \right|.
\end{eqnarray*}
Therefore, under the bounded assumption, the sequence $\log \left|\int_E f^{(n)}(w)\bx(w)\bx^{T}(w)dw \right|$ is uniformly bounded and increasing, and hence it is convergent.

Using Lemma 1 and Lemma 2, we can obtain
\begin{eqnarray*}
0 & = & \lim_{n \to \infty}\log\left|\int_E f^{(n)}(w)\bx(w)\bx^{T}(w)dw \right| - \log\left|\int_E f^{(n-1)}(w)\bx(w)\bx^{T}(w)dw \right|\\
& \geq & p \int_E f^{(n)}(w)\log \frac{f^{(n)}(w)}{f^{(n-1)}(w)}dw\\
& \geq & \frac{p}{2}[\int_E |f^{(n)}(w)- f^{(n-1)}(w)|dw]^{2}\geq 0.
\end{eqnarray*}
Then, we can conclude that
$$
\int_E |f^{(n)}(w)-f^{(n-1)}(w)|dw\longrightarrow 0 {\mbox { as }} n \longrightarrow +\infty. 
$$

\subsection*{Proof of Theorem 3}


It is easy to check that ${\mbox {tr}}(\textbf{A})^{-1}$ is concave in $\textbf{A}$, where $\textbf{A}$ is positive definite matrix. For the sake of simplicity, we make $S = \int_E f(w)\bx(w)\bx^{T}(w)dw$ and $S^{*}=\int_E f^{*}(w)\bx(w)\bx^{T}(w)dw$. We have
\begin{eqnarray*}
{\mbox {tr}}[(1-\lambda)S^{*}+\lambda S]^{-1} \leq (1-\lambda){\mbox {tr}}(S^{*-1})+\lambda {\mbox {tr}}(S^{-1}).
\end{eqnarray*}
Then $f^{*}$ is the optimal solution for the $D$-optimal criterion in (2) if and only if for all $f(w)(f(w)\geq 0\;\mbox{and}\;\int_E f(w) dw=1)$,
\begin{eqnarray*}
\frac{{\mbox {tr}}[(1-\lambda)S^{*}+\lambda S]^{-1} - {\mbox {tr}}(S^{*-1})}{\lambda} \geq 0
\end{eqnarray*}
for $\lambda>0$.
Thus, for $\lambda\downarrow 0$ we have
\begin{eqnarray*}
\lim\limits_{\lambda\downarrow 0} \frac{{\mbox {tr}}[(1-\lambda)S^{*}+\lambda S]^{-1}-{\mbox {tr}}(S^{*-1})}{\lambda} & = & \frac{\partial \{ tr[(1-\lambda)S^{*}+\lambda S]^{-1}\}}{\partial \lambda}|_{\lambda=0}\\
& = & -{\mbox {tr}}[S^{*-1}(S-S^{*})S^{*-1} ]\\
& = & -{\mbox {tr}}( S^{*-1}SS^{*-1}) + {\mbox {tr}}( S^{*-1}) \geq 0,
\end{eqnarray*}
which implies  Theorem 3. 

\section*{References}
\begin{description}
\item Anderson, T. W., 2003. An introduction to multivariate statistical analysis, Wiley, New York.
\item Atkinson, A. C., Donev, A.N. and Tobias, R. D., 2007. Optimum experimental designs, With SAS. Oxford University Press.
\item Barker, T. B., 1994. Quality by experimental design. New York, Dekker.
\item Castro, Y. D., Gamboa, F., Henrion, D., Hess, R., and Lasserre, J. - B., 2018. Approximate optimal designs for multivariate polynomial regression. Submited to Annals of Statistics.
\item Chen, Y. H., 2003. D-optimal designs for linear and quadratic polynomial models. Taiwan: National Sun Yat-Sen University.
\item Dennis, D. and Meredith, J., 2000. An empirical analysis of process industry transformation system. Management Science, {\bf 46}(8),1085--1099
\item Dette, H. and Studden, W. J., 1997. The theory of canonical moments with applications in statistics, probability, and analysis. Wiley \& Sons.
\item Fedorov, V., 1972. Theory of optimal experiments, Academic Press, New York.
\item Fransoo, J. C., and Rutten, W. G. M. M., 1994. A Typology of Production control situation in process industries. International Journal of Operations $\&$ Production Management, {\bf 14}(12),47--57.
\item Gao, W., Chan, P. S., Ng, H. K. T. and Lu, X., 2014. Efficient computational algorithm for optimal allocation in regression models. Journal of Computational and Applied Mathematics. {\bf 261}(4), 118--126.
\item Haines, L. M., 1987. The application of the annealing algorithm to the construction of exact optimal designs for linear-regression models. Technometrics, {\bf 29}(4), 439--447.
\item Harman, R. and Filov{\'a}, L., 2014. Computing efficient exact designs of experiments using integer quadratic programming. Computational Statistics \& Data Analysis, {\bf 71}(1), 1159--1167.
\item Harman, R., Filov{\'a}, L. and Richt{\'a}rik, P., 2018. A Randomized Exchange Algorithm for Computing Optimal Approximate Designs of Experiments. https://arxiv.org/abs /1801.05661.
\item Kullback, S., 1967. A lower bound for discrimination information terms of variation, IEEE Transactions on Information Theory, \textbf{13}(1), 126--127.
\item Meyer, R. K. and Nachtsheim, C. J., 1995. The coordinate-exchange algorithm for constructing exact optimal experimental designs, Technometrics, {\bf 37}(1), 60¨C-69.
\item Silvey, S. D., Titterington, D. M., and Torsney, B., 1978. An algorithm for optimal designs on a finite design space, Commun. Stat. Theory Methods, 16(14), 1379--1389.
\item Uci{\'n}ski, D. and Patan, M., 2007. D-optimal design of a monitoring network for parameter estimation of distributed systems. Journal of Global Optimization, {\bf 39}(2), 291--322.
\item Vanhatalo, E., V\"{a}nnman, K. and Hyllander, G., 2007. A designed experiment in a continuous process. Proceedings from the 15th QMOD Quality and Service Sciences, Poznan, Poland.
\item Vanhatalo, E. and Bergquist, B., 2007. Special considerations when planning experiments in a continuous process.
Quality Engineering, {\bf 19}, 155--169.
\item Welch, W. J., 1982. Branch-and-bound search for experimental designs based on D-optimality and other criteria. Technometrics, {\bf 24}(1), 41--48.
\item Wynn, H. P., 1970. The sequential generation of D-optimum experimental designs. Annals of Mathematical Statistics, {\bf 41}(5), 1655--1664.
\item Yu, Y., 2011. D-optimal designs via a cocktail algorithm. Statistics and Computing, {\bf 21}(4), 475--481
\end{description}

\newpage

\begin{figure}[H]
  \centering
  \includegraphics[width=0.9\textwidth]{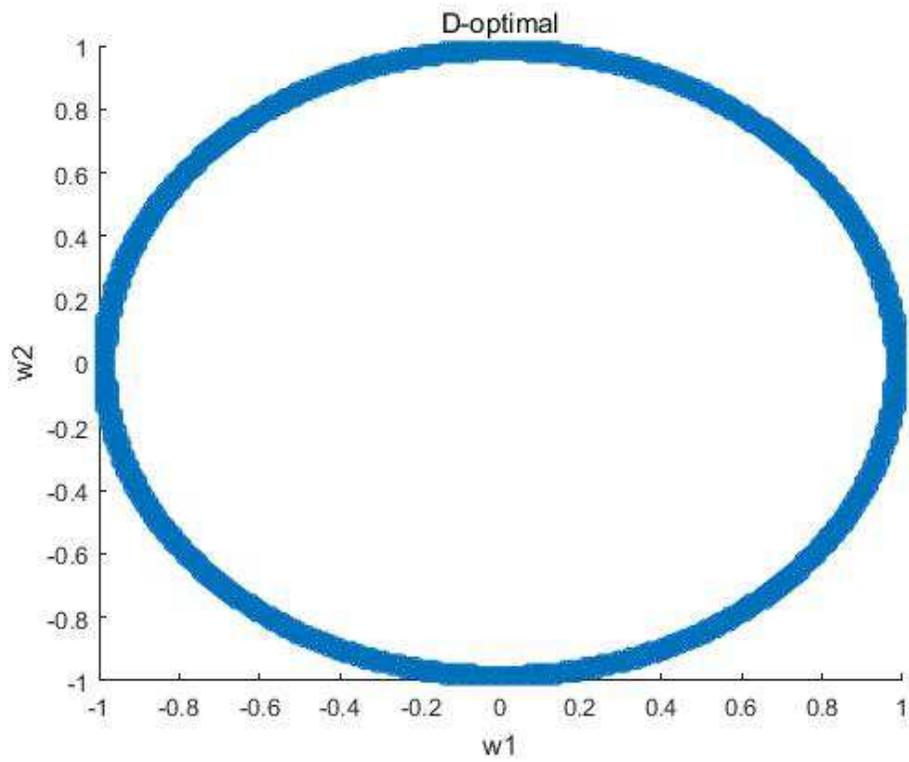}\\
  \caption{$D$-optimal design for Setting 1}\label{1}
\end{figure}

\begin{figure}[H]
  \centering
  \includegraphics[width=0.9\textwidth]{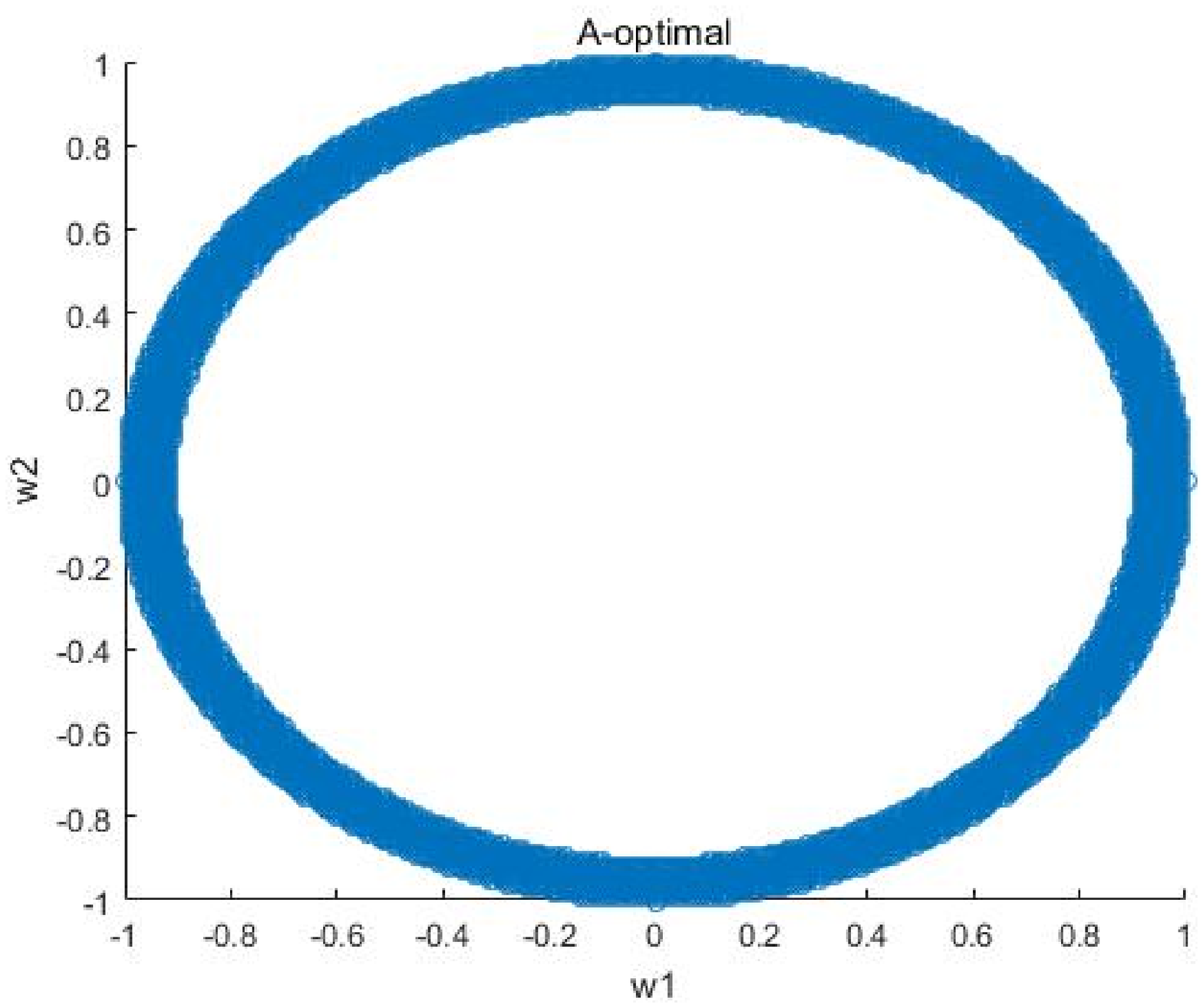}\\
  \caption{$A$-optimal design for Setting 1}\label{2}
\end{figure}

\begin{figure}[H]
  \centering
  \includegraphics[width=0.9\textwidth]{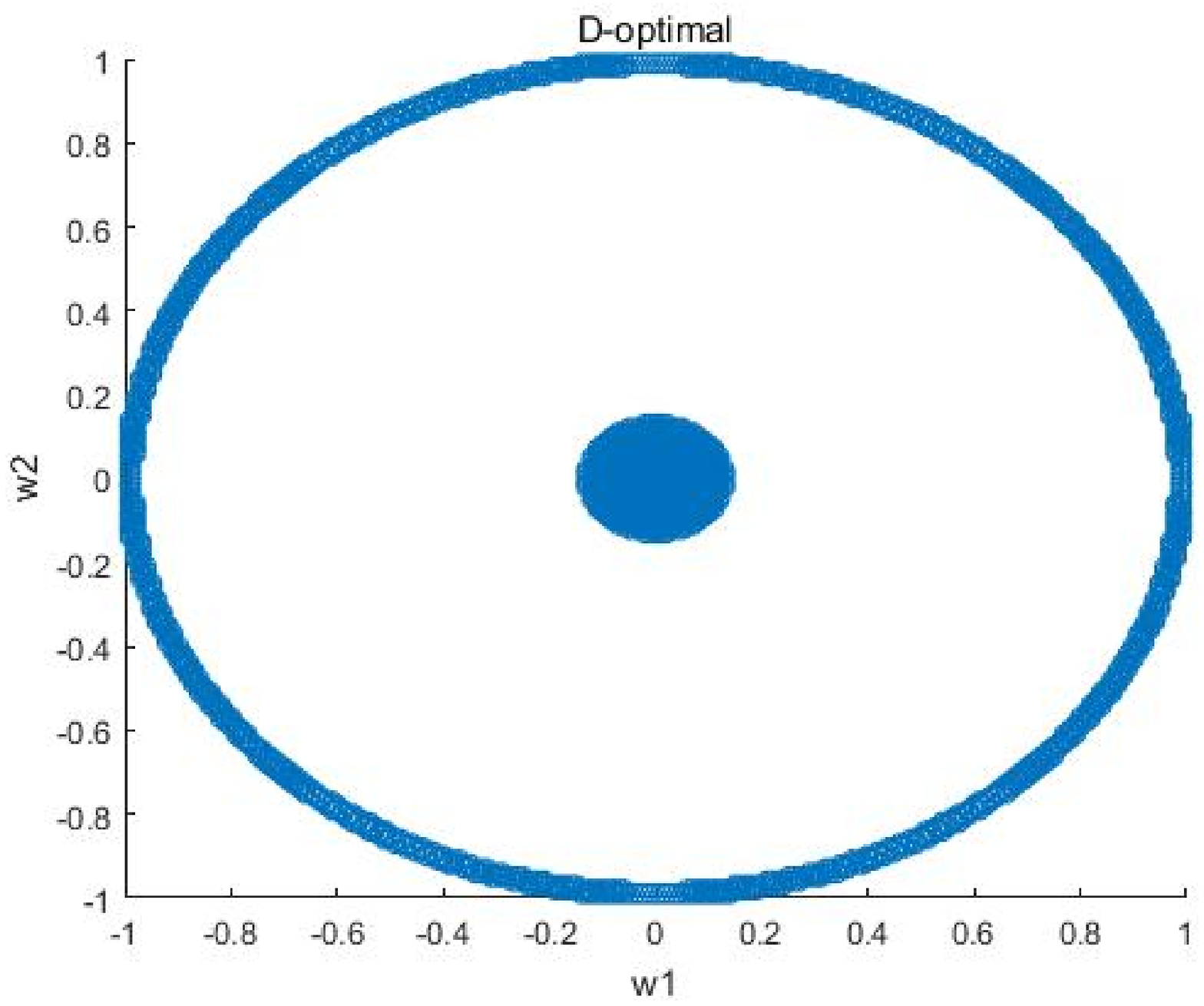}\\
  \caption{$D$-optimal design for Setting 2}\label{3}
\end{figure}

\begin{figure}[H]
  \centering
  \includegraphics[width=0.9\textwidth]{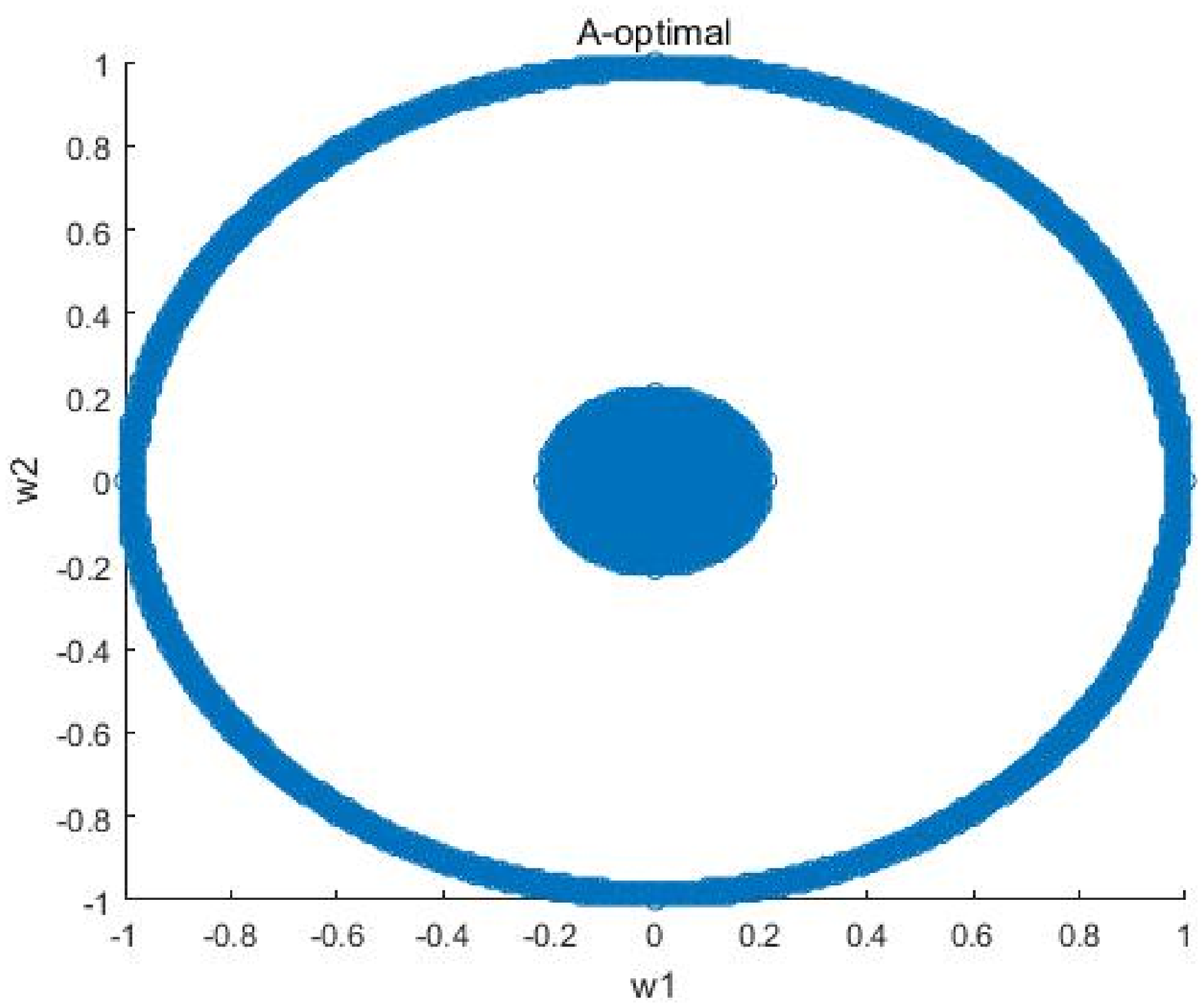}\\
  \caption{$A$-optimal design for Setting 2}\label{4}
\end{figure}

\begin{figure}[H]
  \centering
  \includegraphics[width=0.9\textwidth]{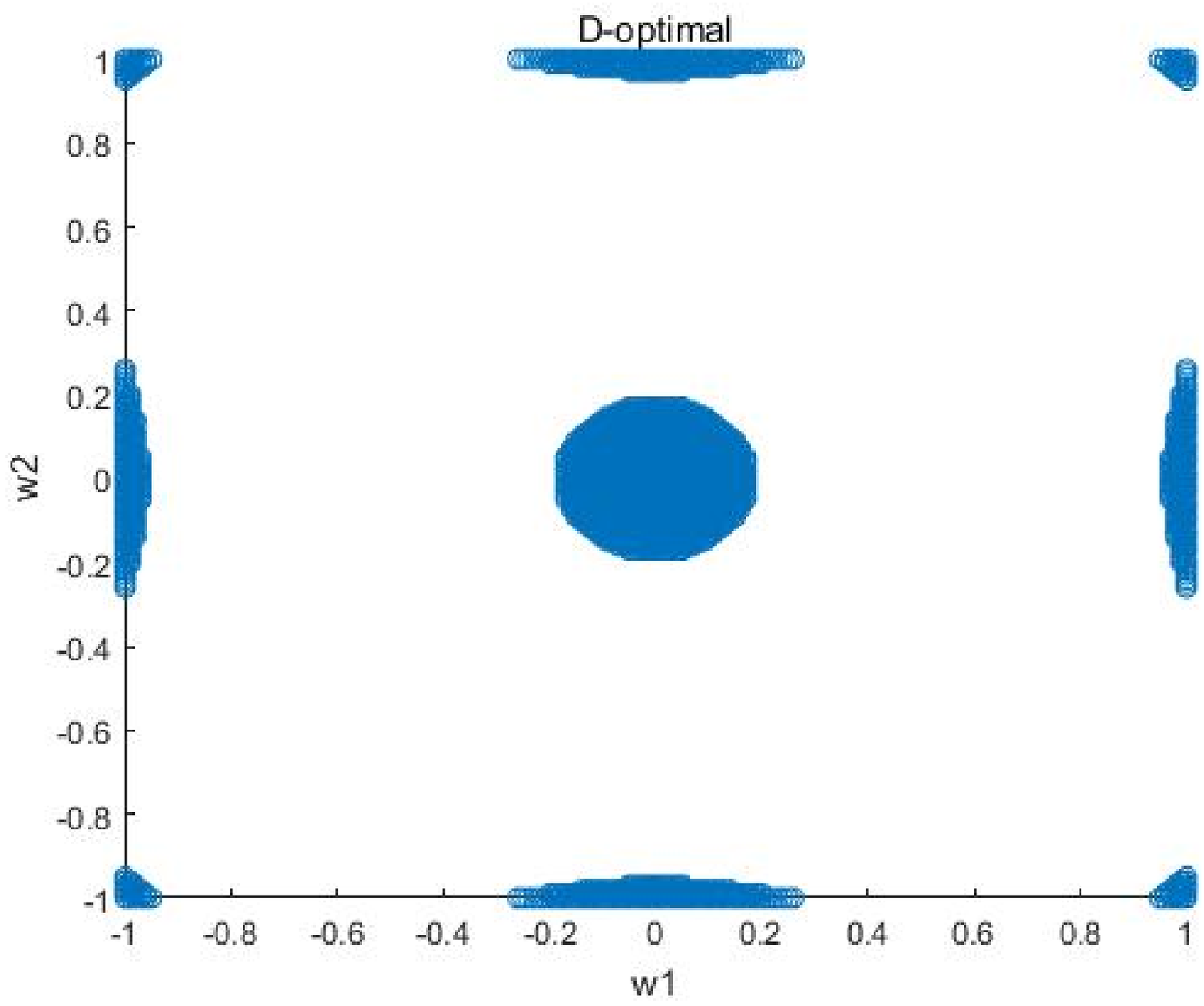}\\
  \caption{$D$-optimal design for Setting 3}\label{1}
\end{figure}

\begin{figure}[H]
  \centering
  \includegraphics[width=0.9\textwidth]{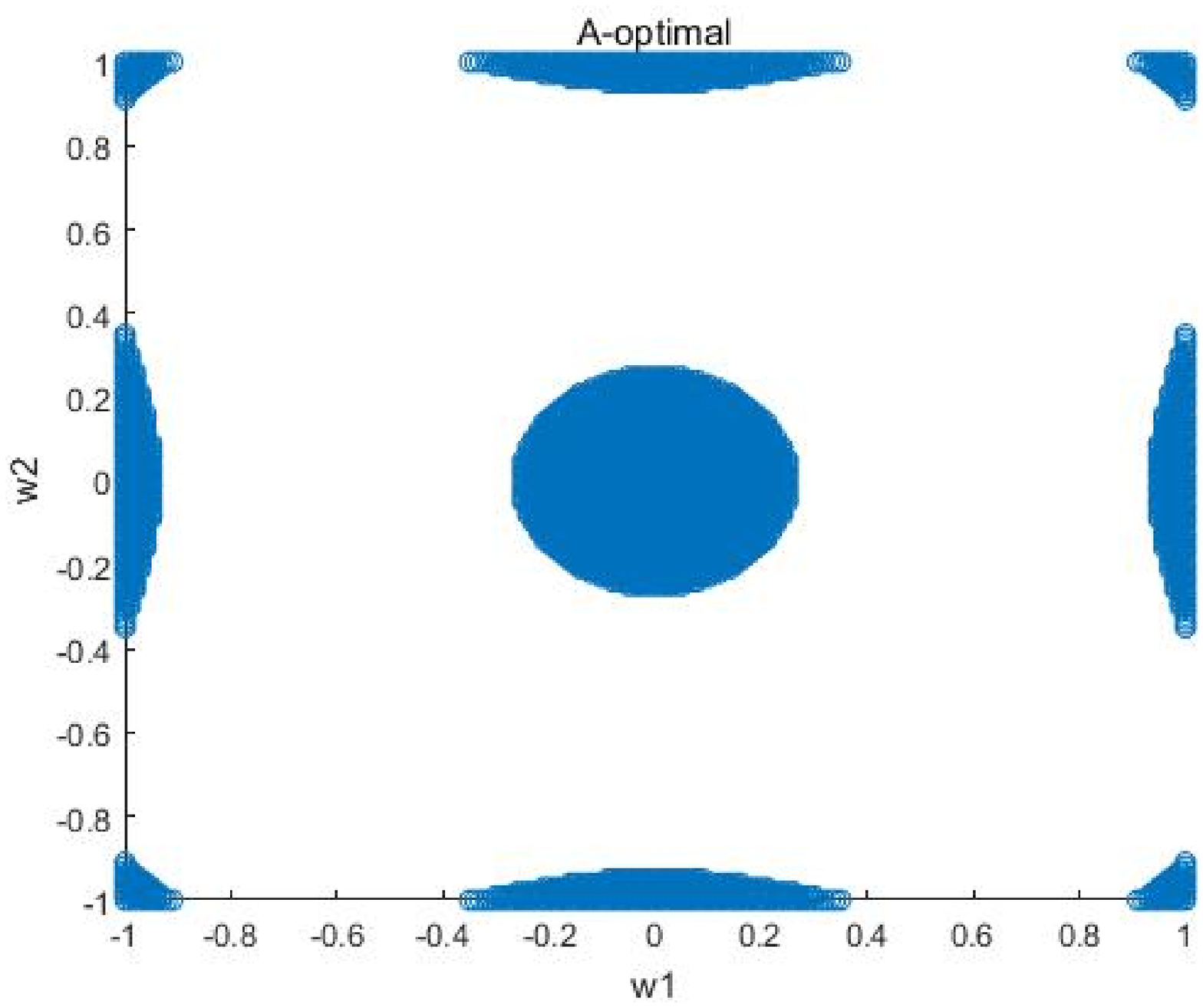}\\
  \caption{$A$-optimal design for Setting 3}\label{1}
\end{figure}

\begin{figure}[H]
  \centering
  \includegraphics[width=0.9\textwidth]{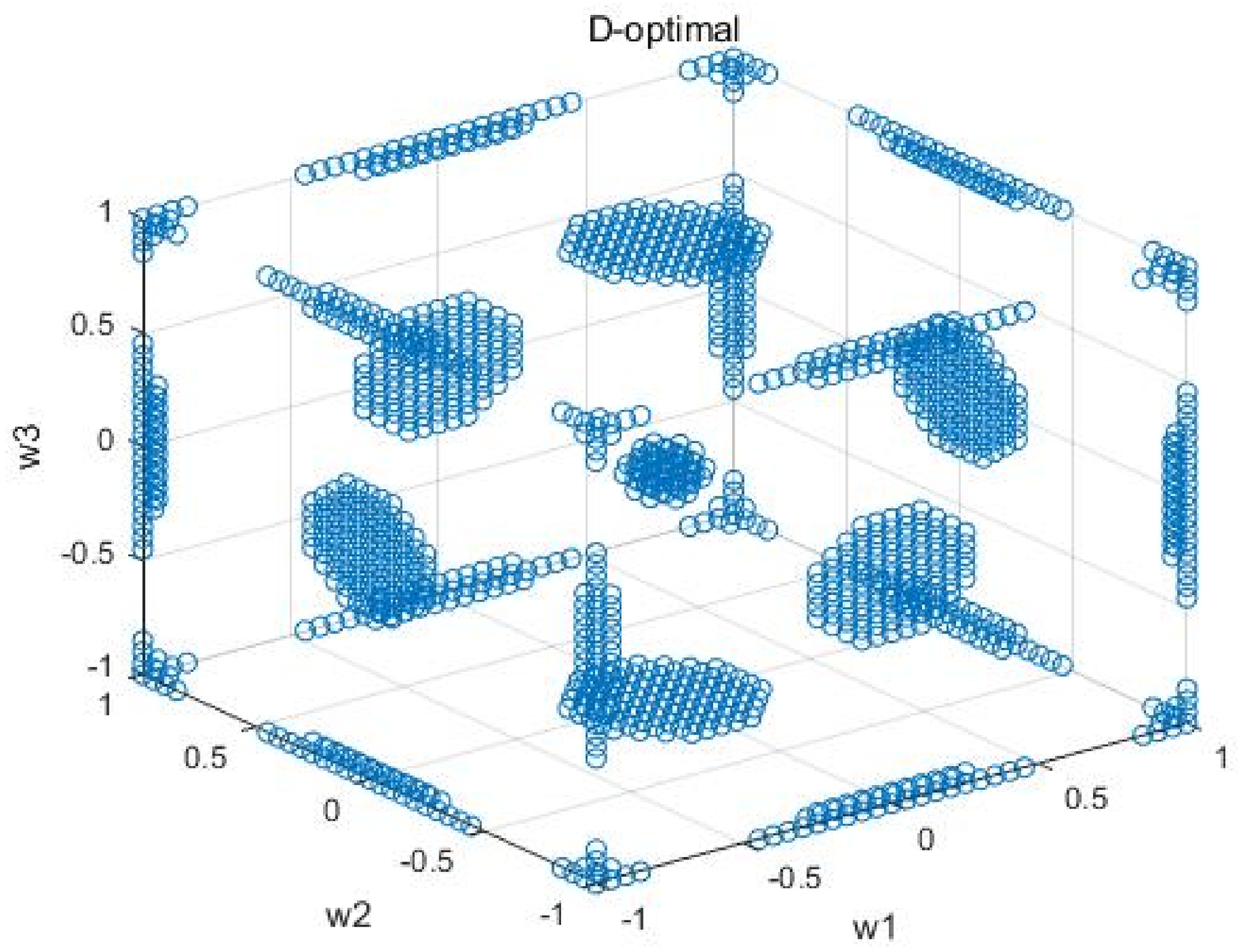}\\
  \caption{$D$-optimal design for Setting 4}\label{1}
\end{figure}

\begin{figure}[H]
  \centering
  \includegraphics[width=0.9\textwidth]{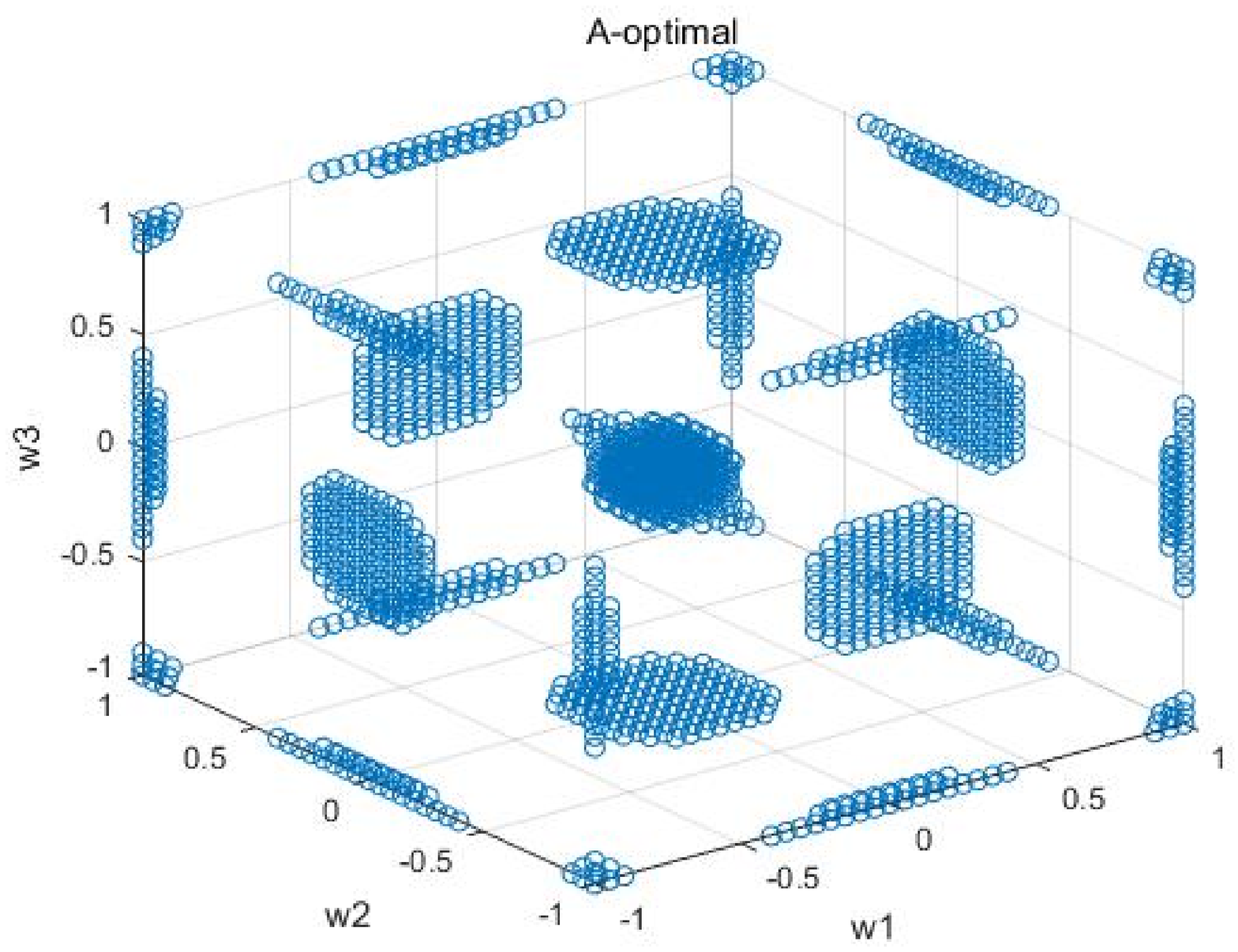}\\
  \caption{$A$-optimal design for Setting 4}\label{1}
\end{figure}

\begin{table}[H]
\newcommand{\tabincell}[2]{\begin{tabular}{@{}#1@{}}#2\end{tabular}}
 \renewcommand{\tabcolsep}{0.2pc}
\renewcommand{\arraystretch}{0.6}
  \centering
  \caption{$D$-optimal design points and weights obtained by the randomized exchange algorithm (REX) for Setting 2}
  \begin{tabular}{c c}\hline
Design points & Weights (from REX)  \\  \hline
$(-0.0549, -0.9985)$ &  0.0682 \\
$(-0.7841, 0.6204)$  &  0.0614 \\
$(0.3688, -0.9294)$   &  0.0840 \\
$(0.0132, 0.9998)$    &  0.1370 \\
$(0.9249, -0.3800)$  &  0.1153 \\
$(0.8617, 0.5074)$    &  0.1306 \\
$(-0.7789, -0.6271)$ &  0.1343 \\
$(0.0000, 0.0000)$    &  0.1667 \\
$(-0.9564, 0.2917)$  &  0.0978 \\
$(-0.3948, 0.9186)$  &  0.0047 \\ \hline
\end{tabular}
\label{DS2}
\end{table}

\begin{table}[H]
\newcommand{\tabincell}[2]{\begin{tabular}{@{}#1@{}}#2\end{tabular}}
 \renewcommand{\tabcolsep}{0.2pc}
\renewcommand{\arraystretch}{0.6}
  \centering
  \caption{$A$-optimal design points and weights obtained by the randomized exchange algorithm (REX) for Setting 2}
  \begin{tabular}{c c}\hline
Design points & Weights (from REX)  \\  \hline
$(-0.0549, -0.9985)$ &  0.0371 \\
$(-0.7841, 0.6204)$  &  0.0355 \\
$(0.3688, -0.9294)$  & 0.0861 \\
$(0.0132, 0.9998)$   & 0.1053\\
$(-0.8859, 0.4637)$  & 0.0978\\
$(0.8109, 0.5851)$   & 0.0781\\
$(0.9249, -0.3800)$  &0.0984\\
$(0.8617, 0.5074)$   & 0.0374\\
$(-0.7789, -0.6271)$ & 0.1324\\
$(0.0004, -0.0010)$  & 0.2919\\ \hline
\end{tabular}
\label{AS2}
\end{table}

\begin{table}[H]
\newcommand{\tabincell}[2]{\begin{tabular}{@{}#1@{}}#2\end{tabular}}
 \renewcommand{\tabcolsep}{0.2pc}
\renewcommand{\arraystretch}{0.6}
  \centering
   \caption{$D$-optimal design points and weights obtained by the randomized exchange algorithm (REX) and the proposed algorithm for Setting 3}
\begin{tabular}{c c c c} \hline
              &  Weights    & Weights   & Weights  \\
Design points & (from REX)  & (from proposed algorithm) &  (theoretical values)  \\  \hline
$(-1, -1)$ & 0.1458 & 0.1457 & 0.1457\\
$(1, 1)$   & 0.1458 & 0.1457 & 0.1457\\
$(-1, 1)$  & 0.1458 & 0.1457 & 0.1457\\
$(1, -1)$  & 0.1458 & 0.1457 & 0.1457\\
$(0, -1)$  & 0.0802 & 0.0804 & 0.0803\\
$(0, 1)$   & 0.0802 & 0.0804 & 0.0803\\
$(-1, 0)$  & 0.0802 & 0.0804 & 0.0803\\
$(1, 0)$   & 0.0802 & 0.0804 & 0.0803\\
$(0, 0)$   & 0.0962 & 0.0955 & 0.0960\\ \hline
\end{tabular}
\label{DS3}
\end{table}

\begin{table}[H]
\newcommand{\tabincell}[2]{\begin{tabular}{@{}#1@{}}#2\end{tabular}}
 \renewcommand{\tabcolsep}{0.2pc}
\renewcommand{\arraystretch}{0.6}
  \centering
   \caption{$A$-optimal design points and weights obtained by the randomized exchange algorithm (REX) and the proposed algorithm for Setting 3}
\begin{tabular}{c c c c} \hline
              &  Weights    & Weights   & Weights  \\
Design points & (from REX)  & (from proposed algorithm) &  (theoretical values)  \\  \hline
$(-1, -1)$ & 0.0940 & 0.0956 & 0.0940\\
$(1, 1)$   & 0.0940 & 0.0956 & 0.0940\\
$(-1, 1)$  & 0.0940 & 0.0956 & 0.0940\\
$(1, -1)$  & 0.0940 & 0.0956 & 0.0940\\
$(0, -1)$  & 0.0978 & 0.0990 & 0.0978\\
$(0, 1)$   & 0.0978 & 0.0990 & 0.0978\\
$(-1, 0)$  & 0.0978 & 0.0990 & 0.0978\\
$(1, 0)$   & 0.0978 & 0.0990 & 0.0978\\
$(0, 0)$   & 0.2332 & 0.2216 & 0.2332\\ \hline
\end{tabular}
\label{AS3}
\end{table}

\begin{table}[H]
\newcommand{\tabincell}[2]{\begin{tabular}{@{}#1@{}}#2\end{tabular}}
 \renewcommand{\tabcolsep}{0.2pc}
\renewcommand{\arraystretch}{0.6}
  \centering
   \caption{$D$-optimal design points and weights obtained by the randomized exchange algorithm (REX) and the proposed algorithm for Setting 4}

	\begin{tabular}{c c c c}	\hline
              &  Weights    & Weights    & Weights\\
Design points & (from REX)  & (from proposed algorithm) &  (theoretical values)  \\  \hline
$(-1, -1, -1)$ &0.0668&0.0685&0.0684\\
$(-1, -1, 1)$ &0.0810&0.0685&0.0684\\
$(-1, 1, -1)$ &0.0625&0.0685&0.0684\\
$(-1, 1, 1)$ &0.0766&0.0685&0.0684\\
$(1, -1, -1)$ &0.0673&0.0685&0.0684\\
$(1, -1, 1)$ &0.0815&0.0685&0.0684\\
$(1, 1, -1)$ &0.0630&0.0685&0.0684\\
$(1, 1, 1)$ &0.0771&0.0685&0.0684\\
$(-1, -1, 0)$ &0.0151&0.0281&0.0262\\
$(-1, 0, -1)$ &0.0336&0.0281&0.0262\\
$(-1, 0, 1)$ &0.0053&0.0281&0.0262\\
$(-1, 1, 0)$ &0.0238&0.0281&0.0262\\
$(0, -1, -1)$ &0.0288&0.0281&0.0262\\
$(0, -1, 1)$ &0.0005&0.0281&0.0262\\
$(0, 1, -1)$ &0.0374&0.0281&0.0262\\
$(0, 1, 1)$ &0.0091&0.0281&0.0262\\
$(1, -1, 0)$ &0.0141&0.0281&0.0262\\
$(1, 0, -1)$ &0.0326&0.0281&0.0262\\
$(1, 1, 0)$ &0.0228&0.0281&0.0262\\
$(1, 0, 1)$ &0.0043&0.0281&0.0262\\
$(-1, 0, 0)$ &0.0318&0.0151&0.0183\\
$(0, -1, 0)$ &0.0414&0.0151&0.0183\\
$(0, 0, -1)$ &0.0045&0.0151&0.0183\\
$(0, 0, 1)$ &0.0611&0.0151&0.0183\\
$(0, 1, 0)$ &0.0241&0.0151&0.0183\\
$(1, 0, 0)$ &0.0339&0.0151&0.0183\\
$(0, 0, 0)$ &0.0000&0.0245&0.0290\\ \hline
	\end{tabular}
\label{DS4}
\end{table}

\begin{table}[H]
\newcommand{\tabincell}[2]{\begin{tabular}{@{}#1@{}}#2\end{tabular}}
 \renewcommand{\tabcolsep}{0.2pc}
\renewcommand{\arraystretch}{0.6}
  \centering
   \caption{$A$-optimal design points and weights obtained by the randomized exchange algorithm (REX) and the proposed algorithm for Setting 4}

	\begin{tabular}{c c c c}	\hline
              &  Weights    & Weights    & Weights\\
Design points & (from REX)  & (from proposed algorithm) &  (theoretical values)  \\  \hline
$(-1, -1, -1)$ & 0.0277 &0.0373&0.0402\\
$(-1, -1, 1)$ &0.0264 &0.0373&0.0402\\
$(-1, 1, -1)$ &0.0364&0.0373&0.0402\\
$(-1, 1, 1)$ &0.0351&0.0373&0.0402\\
$(1, -1, -1)$ &0.0450&0.0373&0.0402\\
$(1, -1, 1)$ &0.0438&0.0373&0.0402\\
$(1, 1, -1)$ &0.0538&0.0373&0.0402\\
$(1, 1, 1)$ &0.0525&0.0373&0.0402\\
$(-1, -1, 0)$ &0.0521&0.0313&0.0259\\
$(-1, 0, -1)$ &0.0421&0.0313&0.0259\\
$(-1, 0, 1)$ &0.0447&0.0313&0.0259\\
$(-1, 1, 0)$ &0.0347&0.0313&0.0259\\
$(0, -1, -1)$ &0.0335&0.0313&0.0259\\
$(0, -1, 1)$ &0.0361&0.0313&0.0259\\
$(0, 1, -1)$ &0.0161&0.0313&0.0259\\
$(0, 1, 1)$ &0.0186&0.0313&0.0259\\
$(1, -1, 0)$ &0.0175&0.0313&0.0259\\
$(1, 0, -1)$ &0.0075&0.0313&0.0259\\
$(1, 1, 0)$ & 0.0000&0.0313&0.0259\\
$(1, 0, 1)$ &0.0100&0.0313&0.0259\\
$(-1, 0, 0)$ &0.0080&0.0330&0.0430\\
$(0, -1, 0)$ &0.0253&0.0330&0.0430\\
$(0, 0, -1)$ &0.0453&0.0330&0.0430\\
$(0, 0, 1)$ &0.0405&0.0330&0.0430\\
$(0, 1, 0)$ &0.0602&0.0330&0.0430\\
$(1, 0, 0)$ &0.0774&0.0330&0.0430\\
$(0, 0, 0)$ &0.1101&0.1288&0.1096\\ \hline
	\end{tabular}
\label{AS4}
\end{table}

\end{document}